\documentclass[runningheads]{llncs}
\usepackage{amsmath}
\usepackage[T1]{fontenc}
\usepackage{graphicx}
\begin{document}
\title{Fine-Tuning Large Language Models Using EEG Microstate Features for Mental Workload Assessment  \thanks{Supported by: School of Computer Science, Technological University Dublin}}
%
%
\author{Bujar Raufi\inst{1,2}\orcidID{0000-0002-0153-6144} }
\authorrunning{B. Raufi.}
%
\institute{School of Computer Science, Technological University Dublin, Dublin, Ireland
\email{bujar.raufi@tudublin.ie}\\
\url{https://www.tudublin.ie/explore/faculties-and-schools/computing-digital-data/school-of-computer-science/} \and Artificial Intelligence and Cognitive Load Lab, Applied Intelligence Research Centre, School of Computer Science, Technological University Dublin, Dublin, Ireland
}
\maketitle              
\begin{abstract}
This study explores the intersection of electroencephalography (EEG) microstates and Large Language Models (LLMs) to enhance the assessment of cognitive load states. By utilizing EEG microstate features, the research aims to fine-tune LLMs for improved predictions of distinct cognitive states, specifically 'Rest' and 'Load'. The experimental design is delineated in four comprehensive stages: dataset collection and preprocessing, microstate segmentation and EEG backfitting, feature extraction paired with prompt engineering, and meticulous LLM model selection and refinement. Employing a supervised learning paradigm, the LLM is trained to identify cognitive load states based on EEG microstate features integrated into prompts, producing accurate discrimination of cognitive load. A curated dataset, linking EEG features to specified cognitive load conditions, underpins the experimental framework. The results indicate a significant improvement in model performance following the proposed fine-tuning, showcasing the potential of EEG-informed LLMs in cognitive neuroscience and cognitive AI applications. This approach not only contributes to the understanding of brain dynamics but also paves the way for advancements in machine learning techniques applicable to cognitive load and cognitive AI research.
\end{abstract}

\section{Introduction}
\label{sec:intro}

Large Language Models (LLMs) have revolutionized the field of natural language processing (NLP) by demonstrating remarkable capabilities across various tasks. These models, which are typically based on transformer architectures and trained on large amounts of data, have shown significant improvements in performance as their size increases. LLMs exhibit some cognitive abilities similar to humans, particularly in formal linguistic tasks and certain reasoning scenarios\cite{Zhuang2023Efficiently,Mahowald2023Dissociating,Momennejad2023Evaluating}. However, they fail in more complex cognitive tasks requiring deeper understanding and planning. While advanced LLM models like GPT-4 or Llama 3.2 show improvements, significant differences remain between LLMs and human cognitive systems\cite{Sartori2023Language,Goertzel2023Generative}.

A promising avenue for enhancing the cognitive capabilities of AI systems lies in integrating biological data that reflect the underlying cognitive processes. Among the various neurophysiological measures, electroencephalography (EEG) microstates have emerged as significant markers of cognitive function. EEG microstates represent transient, patterned, quasi-stable states of EEG brain activity. These quasi-stable brain activities in a range of milliseconds are thought to reflect the temporal dynamics of neural processing involved in perception, attention, and information integration, thus often nicknamed the "atoms of thought"~\cite{koenig2002millisecond}.

EEG microstates have been associated with specific cognitive processes. Changes in microstate parameters (duration, occurrence, and coverage) are influenced by cognitive tasks, indicating their role in cognitive processing \cite{Chen2023Evidence,Milz2016The,Seitzman2017Cognitive}. EEG microstates correlate with resting-state networks identified by fMRI, suggesting that they reflect the activity of large-scale brain networks \cite{Michel2017EEG,Khanna2015Microstates,Lian2021Altered}. The temporal dynamics of EEG microstates are altered in various cognitive and mental states, including neurological and psychiatric conditions, indicating their significance in cognitive processes \cite{Michel2017EEG,Khanna2015Microstates,Ville2010EEG}. EEG microstates are influenced by mental workload and task types, with different tasks affecting the parameters and topographies of these microstates in varied ways. Specific microstates are sensitive to cognitive manipulations, such as attention tasks and visual input, further supporting their role in cognitive functions \cite{Seitzman2017Cognitive}.
The microstates of the EEG change during altered states of consciousness, such as sleep, anaesthesia, and meditation, suggesting their involvement in the underlying characteristics of self-consciousness \cite{Bréchet2022EEG}. Alterations in EEG microstate dynamics are observed in conditions such as mild cognitive impairment (MCI) and Alzheimer's disease, indicating their potential as biomarkers for cognitive impairment \cite{Lian2021Altered}.
Differences in microstate parameters are also found in children and adolescents with autism spectrum development, reflecting the atypical activity of the resting state network \cite{Takarae2022EEG}. In general, EEG microstates play a significant role in cognitive processes, reflecting the functional state of the brain and its large-scale network dynamics. They are influenced by cognitive tasks, mental workload, and altered states of consciousness and show potential as biomarkers of cognitive impairments and developmental conditions. The study of EEG microstates provides valuable information on the temporal dynamics of brain activity and their association with various cognitive functions.

This paper provides a detailed experiment design on how a potential fine-tuning of LLMs with EEG microstates can be utilized to asses two distinct cognitive load states ("Rest" and "Load"). The experiment design offers sufficient details to allow valid reproducibility. Furthermore, the experiment opens an exciting path to LLM contextualization through fine-tuning with cognitive data reflecting the underlying brain cognitive activities. 

The rest of the paper is organized as follows: section \ref{sec:related} provides an overview of LLMs and their applications in cognitive tasks and discusses the research on EEG microstates and their relationship to cognitive processes as well as tackles the concept of fine-tuning LLMs and its potential benefits. Section \ref{sec:exp_design} provides an experiment design of fine-tuning LLMs with EEG microstates. Section \ref{sec:conclusion} concludes the paper and outlines the future work. 

\section{Related Work}
\label{sec:related}
Large Language Models (LLMs) have emerged as powerful tools in natural language processing, demonstrating capabilities that extend beyond their initial design of predicting the next word in a sequence. These models, such as GPT-4 and others, have shown remarkable performance in various cognitive tasks, often paralleling human-like reasoning and problem-solving abilities \cite{Sartori2023Language,Nolfi2023On,Raiaan2024A}. Large Language Models (LLMs) have shown cognitive abilities that extend beyond their initial training objectives. These capabilities include complex reasoning, problem-solving, and decision-making tasks, which are typically associated with human cognition \cite{Nolfi2023On,Zhu2024Language}. For example, LLMs have been utilized to model human decision-making processes, such as risky and intertemporal choices, by employing computationally equivalent tasks \cite{Zhu2024Language}. Moreover, LLMs can perform zero-shot reasoning, meaning they can solve tasks without having received specific prior examples, indicating a broad cognitive capacity \cite{Kojima2022Large}. Increasingly, researchers in cognitive psychology are leveraging LLMs to investigate the mechanisms underlying intelligence and reasoning. They have been applied to various tasks, including arithmetic, symbolic reasoning, and logical reasoning, often achieving performance levels comparable to human benchmarks \cite{Sartori2023Language,Kojima2022Large}. Moreover, LLMs are used to study cognitive biases in decision-making, providing insights into human-like biases and offering frameworks to mitigate these biases in high-stakes scenarios \cite{Echterhoff2024Cognitive}. Despite their impressive capabilities, LLMs face challenges in specific cognitive tasks. For example, they struggle with planning and understanding complex relational structures, known as cognitive maps, which are essential for goal-directed behaviour. These limitations highlight the need for systematic evaluation protocols, such as CogEval, to better understand and improve LLMs' cognitive abilities \cite{Momennejad2023Evaluating}.

EEG microstates are brief, 60 to 120 milliseconds long time periods during which the brain's electrical activity remains quasi-stable, reflecting the global functional state of the brain. These microstates are thought to represent the basic building blocks of spontaneous conscious mental processes and are associated with large-scale brain networks \cite{Michel2017EEG,Zhu2024Language}. Typically, four microstate classes (A, B, C, and D) are identified, each associated with different cognitive functions and brain networks \cite{Milz2016The,Seitzman2017Cognitive,Michel2017EEG}. Research has shown that EEG microstates are linked to various cognitive processes. For instance, microstate A is often associated with verbal and phonological processing, while microstate B is linked to visual processing \cite{Milz2016The,Croce2020EEG}. Microstate C is related to the default mode network and cognitive control systems, and microstate D is associated with attention reorientation and the dorsal attention network \cite{Seitzman2017Cognitive,Bréchet2022EEG,Zappasodi2019EEG}. Studies seem to indicate that cognitive tasks can alter the spatial and temporal properties of EEG microstates. For example, during tasks requiring attention, such as serial subtraction, the duration and occurrence of microstate D increase, reflecting its association with the dorsal attention network \cite{Seitzman2017Cognitive}. Similarly, visual tasks can increase the occurrence and coverage of microstate B, highlighting its connection to the visual system \cite{Seitzman2017Cognitive,Croce2020EEG}. EEG microstates have also been used to study cognitive impairments, such as mild cognitive impairment (MCI) and Alzheimer's disease (AD). Alterations in microstate dynamics, such as increased duration and coverage of specific microstates, have been observed in these conditions, suggesting changes in brain network functionality \cite{Lian2021Altered,Metzger2023Functional,Musaeus2019Microstates}. For instance, microstate A, linked to the temporal lobes, shows increased occurrence in MCI and AD, which may reflect early pathological changes in these regions \cite{Musaeus2019Microstates}.

Fine-tuning plays an essential role in customizing large language models (LLMs) for specific tasks or domains by modifying the model's parameters using a smaller, dedicated dataset. This method improves the model's effectiveness on specific tasks without requiring the development of an entirely new model. The advantages of fine-tuning include enhanced task performance, increased data efficiency, better domain adaptation, optimized resource use, and greater generalization and versatility.

Large Language Models (LLMs) represent a breakthrough in modelling cognitive tasks, offering insights into human-like reasoning and decision-making. While promising for various applications, ongoing research is essential to address their limitations in AI and cognitive science. EEG microstates provide insights into brain network dynamics linked to cognitive processes. This non-invasive method studies typical cognitive functions and atypical neurological conditions. More research is necessary to understand the relationship between microstates and cognition and their potential applications in AI. Fine-tuning LLMs is one direction and yields many benefits, including enhanced task performance, better data use, and tailored applications, improving their effectiveness across fields. Optimizing resource use and generalization abilities makes LLMs more flexible and efficient for diverse applications.

\section{Experiment Design}
\label{sec:exp_design}
\begin{figure}
    \centering
    \includegraphics[width=1.0\linewidth]{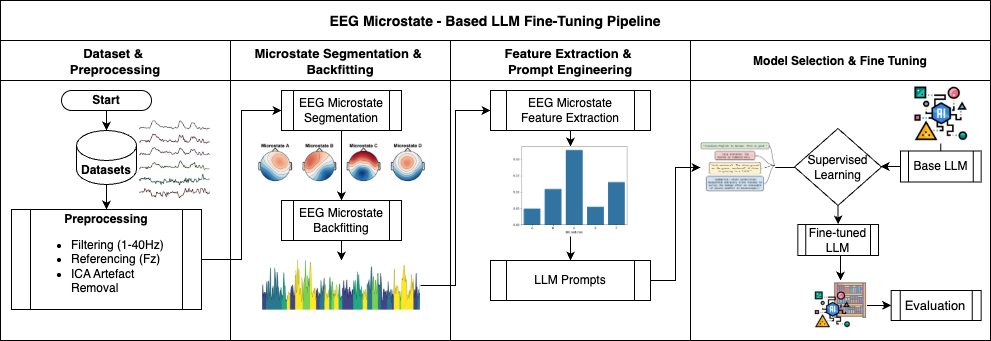}
    \caption{The experiment pipeline of EEG microstate-based LLM fine-tuning.}
    \label{fig:pipeline}
\end{figure}

The proposed experiment design is outlined in four stages: dataset collection and preprocessing, microstate segmentation and EEG backfitting, feature extraction and prompt engineering and LLM model selection and fine-tuning. The overall EEG microstates-based LLM fine-tining experiment pipeline is provided in figure \ref{fig:pipeline}.
\subsection{Dataset Description and Preprocessing}
\label{subsec:dataset_desc}
Two datasets sharing the same task-oriented design are utilised for this LLM fine-tuning. The first dataset is sourced from Zyma et al. \cite{zyma2019electroencephalograms}, while the second dataset consists of data from Shin et al. \cite{shin2016open}. The former utilises data collected from 36 subjects aged between 17 and 29, of whom 27 are female and nine are male. The latter dataset comprises 29 subjects aged between 33 and 44, including 16 males and 17 females. Both datasets introduce a mental arithmetic task involving 4-digit and 3-digit subtraction tasks. The available datasets involving mental arithmetic tasks collected with EEG data are limited, and the rationale for utilising two datasets is twofold: the first is to ensure diverse original datasets that involve mental arithmetic tasks indicating a similar approach in dataset adoption; the second is to obtain EEG microstates that do not differ significantly from the nature of the task.  

Due to the limited number of subjects from which the microstate features were extracted (a total of 103), data synthesis is employed to augment the number of training samples. For this purpose, a Generative Adversarial Network (GAN) was utilised to enhance the training samples. To verify the quality of the synthesised training set, a synthetic quality score was employed to evaluate the goodness of the generated data. Adopting a weighted approach, the Synthetic Data Quality Score is determined by integrating individual quality metrics such as Field Distribution Stability, Field Correlation Stability, and Deep Structure Stability. This score serves as an estimate of the degree to which the synthetic data preserves the statistical properties of the original dataset \cite{raufi2022evaluation}. 

Field Distribution Stability measures how closely the synthetic data distributions match the original data. It uses the Jensen-Shannon Distance to compare numeric or categorical fields. A lower average JS Distance score across fields indicates higher Field Distribution Stability quality. The Field Correlation Stability is calculated by first computing the correlation between each pair of fields in the training data, followed by the synthetic data. The absolute difference between these values is then computed and averaged across all fields. A lower average value indicates a higher quality score for Field Correlation Stability. The deep structure stability is a metric calculated by comparing the distributional distance between the principal components found in the original and synthetic datasets. Gretel AI is used to generate the synthetic training data \footnote{https://gretel.ai/}.

\subsection{EEG Microstate segmentation and backfitting}
\label{subsec:eeg_arch}

A brain microstate is defined by a coordinate vector representing a point at a unit distance from the origin. Any point along the line extending from the origin to this microstate belongs to the same microstate, meaning all such points share a uniquely normalized scalp electric potential field (strictly unique only when an infinite number of electrodes is considered). Consequently, as long as a trajectory remains along this line, the brain remains in the same microstate. The distance from the origin to a point on this line corresponds to the neuronal generators' intensity (or strength) associated with the microstate \cite{pascual1995segmentation}. This distance is also directly proportional to the global field power (GFP’) given  as: 
\begin{equation}
\label{eq:gfp}
    GFP(t) = \sqrt{\frac{1}{N}\sum_{i=1}^{n}(V_i(t)-\overline{V}(t))^2}
\end{equation}
where, $V_i(t)$ represents the electric potential of an electrode $i$ at time $t$, $V(t)$ is the mean potential across all $N$ electrodes at time $t$, and $N$ represents the number of electrodes. The equation \ref{eq:gfp} reflects the instantaneous standard deviation of scalp potential measurements. There is an ongoing debate in the research community whether the microstate identification should focus on GFPs or adopt a wider approach by omitting the GFP and capturing broader brain dynamics \cite{Shaw2019Capturing}. The experiment adopted here followed a well-established methodology outlined by Lehman \cite{lehmann2009eeg} and is a two-step process. 
\begin{enumerate}
    \item In the first step, microstate identification is performed, whereby EEG data is partitioned into clusters using the modified K-means (mod K-means) clustering algorithm. This process involves identifying distinct topographies of electric potentials that remain stable for short periods 
\cite{pascual1995segmentation}. These topographies, often termed archetypes, represent the topography of specific classes and are associated with various cognitive processes such as verbal, visual, and attention-related activities 
\cite{Milz2016The,Michel2017EEG}. 
\item In the second step, the identified microstate archetypes are reinserted into the EEG dataset by substituting the EEG data at each time point with the microstate label that most closely aligns with the EEG topography at that moment\cite{lehmann2009eeg}. This close alignment is quantified Global Explained Variance (GEV) given as:
\begin{equation}
\label{eq:gev}
    GEV = \frac{\sum_{t=1}^{max}(GFP_u(t)\cdot C_{u,T_t})^2 \cdot \gamma_{u,k,t}}{\sum_{t=1}^{max}GFP_u^2(t)}
\end{equation}
where $C_{u,T_t}$ is the spatial correlation between data of a certain condition $U$ at time point $t$, and the respected microstate archetype and the $GFP_u(t)$ is the Global Field Power of EEG signal at the given time $t$. The mapping function $\gamma_{u,k,t}$ represents the data that have been labelled ($L$) as belonging to the $k_th$ segment on EEG signal and is given as:
\begin{equation*}
    \gamma_{u,k,t} = 
        \begin{cases}
            1 & \text{if $k = L_{u,t}$ }\\
            0 & \text{if $k \neq L_{u,t}$}\\
        \end{cases}
\end{equation*}
\end{enumerate}

\subsection{Feature Extraction and Prompt Engineering}
\label{subsec:prompt_eng}

From the back-fitted EEG microstates, five features are extracted. These features are well-established in microstate research and specifically relate to modalities of thinking.
\begin{itemize}
    \item \textbf{Global Explained Variance (gev):} This feature represents the total explained variance expressed by a given state. It is computed as the sum of the global explained variance values of each time point assigned to a given state. Given as a percentage between $0.0-1.0$. 
   \item \textbf{Mean correlation (mean\_corr)} corresponds to the mean correlation value of each time point assigned to a given state. Given as a percentage between $0.0-1.0$
   \item \textbf{Time coverage (timecov)} is the proportion of time during which a given state is active. Given in seconds (s, ms)
   \item \textbf{Mean durations (meandurs)} relates to the mean temporal duration of segments assigned to a given state. Given in seconds or milliseconds.
   \item \textbf{Occurrence per seconds (occurence):} indicates the mean number of segments assigned to a given state per second. This metric is expressed in segments per second and is given in Hertz (Hz).
\end{itemize}

The subsequent step involves crafting the prompts for training the large language model (LLM). This prompt engineering adopts a prompt learning strategy, which utilizes unsupervised prompt learning for classification alongside black-box LLMs. Here, prompts are represented as sequences of discrete tokens featuring learnable categorical distributions. This method takes advantage of the in-context learning abilities of LLMs and integrates pseudo-labeled data as in-context examples throughout the training process \cite{zhang2024unsupervised}. In our case, the learnable categorical distributions are the features embedded within the designed prompts. Finally, the target class represents two distinct prompts outlining the two cognitive load states, the "Rest" state and the "Load" state.

\subsection{LLM Model Selection \& Fine-Tuning}
\label{subsec:fine_tuning}

This study utilizes a pre-trained Large Language Model (LLM) fine-tuned on a dataset consisting of prompts of EEG microstates as mentioned in subsection \ref{subsec:prompt_eng}. Model selection criteria focused on the following considerations:
\begin{itemize}
    \item Models free access, openness and usage.
    \item LLMs that demonstrate strong performance in complex reasoning and sequential data processing, mirroring the structured nature of thought processes related to cognitive load on arithmetic task setting. 
    \item LLM models that excel in tasks like logical reasoning, given their conceptual alignment with arithmetic problem-solving. 
    \item Architectural compatibility with EEG data, publicly available checkpoints, and established fine-tuning procedures essential for reproducibility and comparability with future research.
    \item Model's pre-training data and size to ensure relevance and computational feasibility within the study's scope.
\end{itemize}

The selected LLM undergoes fine-tuning using a curated dataset linking EEG microstate features to corresponding classes of cognitive load states. This dataset encompasses diverse microstate classes across the cognitive state levels to promote generalizability and robustness. We employ a supervised learning approach, training the LLM to predict the cognitive load state based on prompts, including the EEG microstate feature, to produce a response outlining the cognitive load states. The fine-tuning process will optimize model parameters to minimize a suitable loss function, potentially combining cross-entropy loss for solution steps with a distance-based metric for EEG microstate features. Various fine-tuning strategies, including prompt engineering and parameter-efficient methods, will be explored to balance performance gains with computational resource constraints.

\section{Results}
\label{sec:results}
\subsection{Dataset Preprocessing}
\label{subsec:res_data}
The dataset comprises EEG recordings collected from 103 subjects across two sources, as detailed in Section \ref{subsec:dataset_desc}. The first dataset (Dataset 1), provided by Zyma et al. \cite{zyma2019electroencephalograms}, includes discrete EEG segments of 180 seconds for the resting state and 60 seconds for mental counting. The authors of this dataset employed Independent Component Analysis (ICA) to eliminate potential ocular artefacts during the recordings. The second dataset (Dataset 2), derived from Shin et al. \cite{shin2016open}, consists of EEG recordings from subjects engaged in 30-second arithmetic tasks, with intervals of 15 to 20 seconds of rest, repeated across multiple trials. For both datasets, as part of our research methodology, a reference utilizing the 'Fz' channel and bandpass filtering between 1 Hz and 40 Hz, along with ocular artefact removal through ICA, has been applied to the datasets.

\subsection{Microstate Segmentation and Backfitting}
\label{subsec:res_micsostates}
\begin{figure}[!ht]
    \centering
    \includegraphics[width=0.49\linewidth]{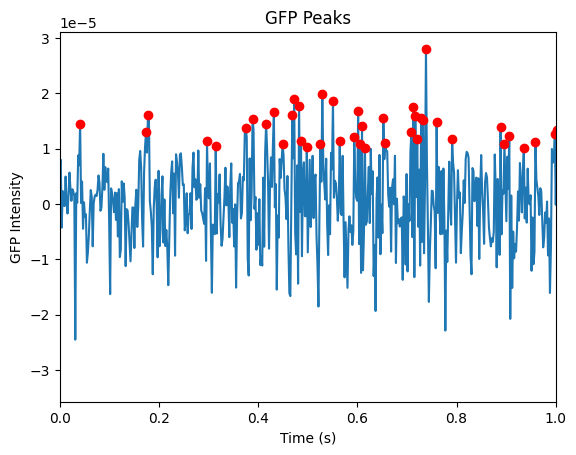}
    \includegraphics[width=0.49\linewidth]{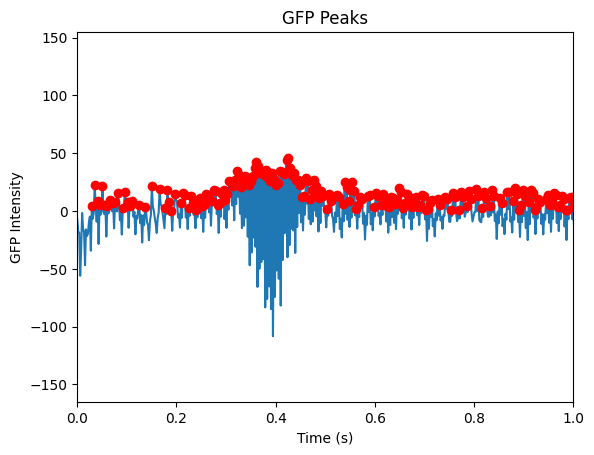}
    \caption{The extracted Global Field Power (GFP) Peaks for Subject 1 in 1-second duration. The image on the left represents the GFps extracted from Dataset 1, and the image on the right from Dataset 2}
    \label{fig:gfps}
\end{figure}

The process of microstate segmentation begins with Global Field Power (GFP) identification as illustrated in subsection \ref{subsec:eeg_arch}. Figure \ref{fig:gfps} illustrates the extracted GFPs for Subject 1 for a duration of 1 second. 
The complete extraction of the Global Field Power (GFP) occurs continuously throughout the entire duration of the electroencephalogram (EEG) recordings during the arithmetic task, which lasts explicitly for 60 seconds for Dataset 1 and 30 seconds for Dataset 2. 
Figure \ref{fig:gfps} illustrates the 1s segment of the GFPs for both datasets for the sake of clarity and brevity. 
\begin{figure}[!ht]
    \centering
    \includegraphics[width=1.0\linewidth]{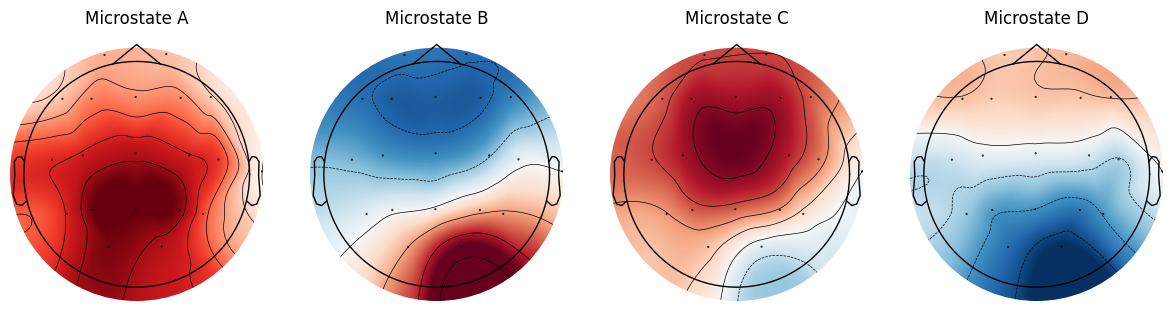}
    \includegraphics[width=1.0\linewidth]{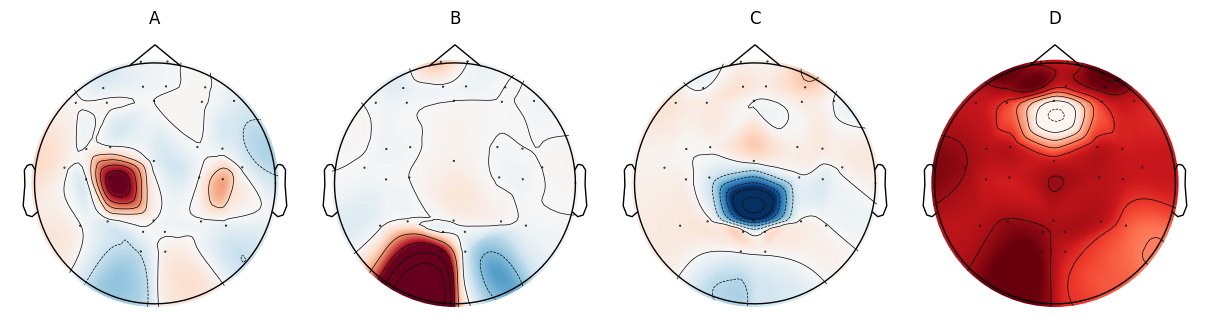}
    \caption{Extracted microstate clusters for Subject 1 from Datasets 1 and 2. The top image illustrates microstates from Dataset 1, while the bottom image displays microstates from Dataset 2.}
    \label{fig:microstates}
\end{figure}
Furthermore, to extract the microstate artefacts (templates), a modified k-means (modKmeans) clustering algorithm was utilized. The algorithm was applied to each subject.  
Figure \ref{fig:microstates} illustrates the extracted four microstate artefacts or templates. 
In the figures depicting microstates, we observe frontal and parietal brain activity clusters indicating cognitive load conditions. These microstates are particularly clear in the second dataset due to the high quality of the EEG data recordings and the sampling rate (1000Hz). Another aspect of the clarity of microstates in the second dataset stems from the authors providing easily extractable EEG recordings of the arithmetic task alone. 

\begin{figure}[!ht]
    \centering
    \includegraphics[width=0.49\linewidth]{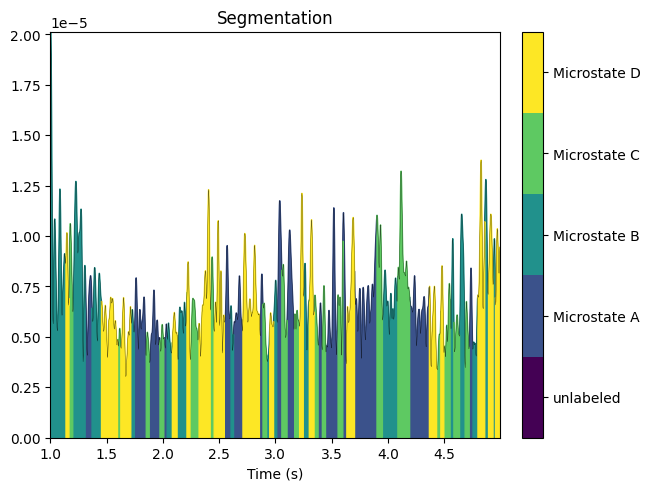}
    \includegraphics[width=0.49\linewidth]{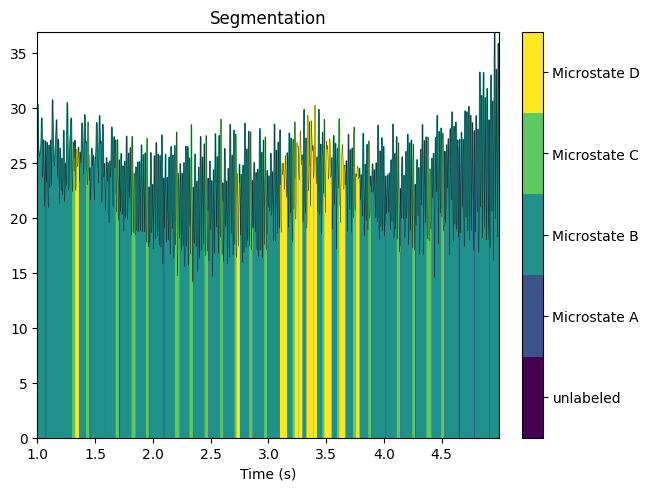}
    \caption{Segmented EEG signal following microstate backfitting. The left image displays the GFPS extracted from Dataset 1, while the right image illustrates the GFPS from Dataset 2.}
    \label{fig:segments}
\end{figure}

Once the microstate clusters have been identified for each of the 103 subjects, the process of backfiiting of these microstates to the EEG signal was applied, resulting in the EEG microstate segmentation. 
\ref{fig:segments} illustrates the segmented EEG signal after microstate backfitting. For the sake of brevity and visibility of the image, only the portion of segmented EEG data is shown for Subject 1. 
The process of microstate cluster extraction and backfitting is carried out across all 103 subjects in both Dataset 1 and Dataset 2. 

\subsection{Feature Selection and Prompt Engineering}
\label{res:fetures_prompts}
Following the analysis of segmented EEG microstates, five distinct features have been extracted: Global Variance ($gev$), Mean Correlation ($mean\_corr$), Time Coverage ($timecov$), Mean Duration ($meandurs$), and Occurrence per Second ($occurrence$). These features have been used to construct prompts based on the following template for prompt generation.
\begin{verbatim}
{ 'user' :<Subject no>,
  {'description': 'Subject of age <age>, a <male, female> 
    with eeg recorded during rest state. Subject's performed a 
    good quality count on number of subtractions achieving a 
    score of <number> during mental arithmetic tasks. Four eeg 
    microstates have been extracted from the subject. Quantitative   
    representation of EEG microstates across five features in a 20 
    minute period have been extracted, the brain activity is 
    segmented into 4 microstates. The feature descriptions used are 
    as follows: <short feature desc...>',
    'query': The following are the parameters for each microstate 
    features:
        Microstate A:
            Global Explained Variance:<gev> seconds. 
            Mean correlation:<mean_corr>. 
            Time coverage:<timecov> seconds. 
            Mean duration <meandurs> seconds. 
            Occurrences:<occurence> times.
        Microstate B:
            Global Explained Variance:<gev> seconds. 
            Mean correlation:<mean_corr>. 
            Time coverage:<timecov> seconds. 
            Mean duration <meandurs> seconds. 
            Occurrences:<occurence> times.              
        Microstate C:
            Global Explained Variance:<gev> seconds. 
            Mean correlation:<mean_corr>. 
            Time coverage:<timecov> seconds. 
            Mean duration <meandurs> seconds. 
            Occurrences:<occurence> times.              
        Microstate D:
            Global Explained Variance:<gev> seconds. 
            Mean correlation:<mean_corr>. 
            Time coverage:<timecov> seconds. 
            Mean duration <meandurs> seconds. 
            Occurrences:<occurence> times.
    Based on the EEG feature parameters above, can you 
    determine the cognitive load state of the subject?', 
    'answer': 'Subject is at <resting, cognitive load> state.'}
}
\end{verbatim}
The initial fine-tuning of the LLM did not yield satisfactory results due to the small number of users and microstate features. The model's accuracy ranged between $50\%$ and $59\%$, and it functioned like a random predictor. To increase the model performance, the training set was expanded with synthetic data, for which the Generative Adversarial Network (GAN) is used. The model parameters used for the GAN model training are illustrated in table \ref{tab:ganparams}.
\begin{table}
    \centering
    \begin{tabular}{|c|c|}
        \hline
        \textbf{GAN Model Parameters} & \textbf{Value} \\ \hline
        Generated Samples & 10,000 \\ \hline
        Batch Size & 1 \\ \hline
        Gradient Accumulation Steps & 8 \\ \hline
        Weight decay & 0.01 \\ \hline
        Warmup ration & 0.05 \\ \hline
        Learning rate scheduler & Cosine \\ \hline
        Learning rate & 0.0005 \\ \hline
        LoRA Ranks  & 32 \\ \hline
        LoRA alpha & 1.0 \\ \hline
        LoRA Target Modules & ["q\_proj", "k\_proj", "v\_proj", "o\_proj"] \\ \hline
    \end{tabular}
    \caption{Placeholder caption for a 2-column, 13-row table.}
    \label{tab:ganparams}
\end{table}
The overall synthetic quality score is $73\%$, indicating a good quality of the synthetic data.
Table \ref{tab:synth_score} outlines the synthetic data quality regarding Field Distribution Stability, Field Correlation Stability, and Deep Structure Stability.
\begin{table}
    \centering
    \begin{tabular}{|c|p{3cm}|p{3cm}|p{3cm}|}
        \hline
        \textbf{Data Instances} & \textbf{Field Distribution Stability} & \textbf{Field Correlation Stability} & \textbf{Deep Structure Stability} \\ \hline
        10,000 & 73\% & 54\% & 93\% \\ \hline
        \textbf{Quality Scale} & Good & Moderate & Excellent \\ \hline
    \end{tabular}
    \caption{Overall synthetic data quality score}
    \label{tab:synth_score}
\end{table}
The table clearly shows that the quality scale for the synthetic data ranges from a moderate correlation stability score to good and excellent distribution and structure stability scores, respectively. To ensure that the generated data closely resembles the original, a correlation analysis is performed between the original and synthetic training datasets. The analysis revealed a strong correlation between the original training set and the synthetic set. 

\begin{figure}[ht]
    \centering
    \includegraphics[width=1.0\linewidth]{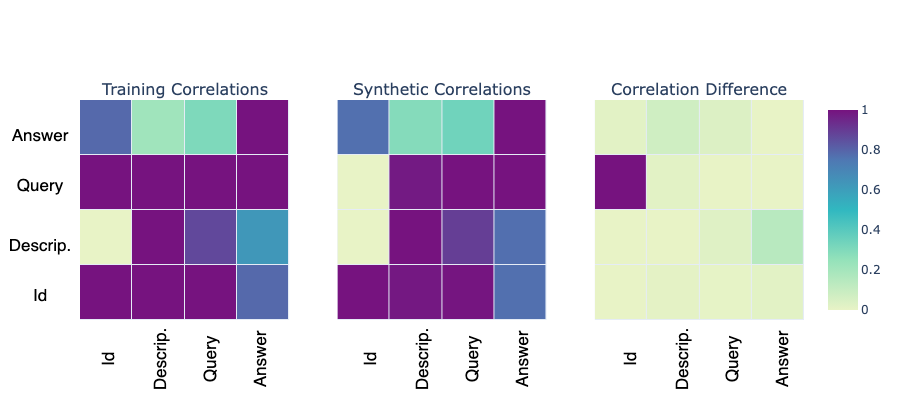}
    \caption{Training and synthetic correlations and correlation difference between original and synthetic data.}
    \label{fig:synth_corr}
\end{figure}

Figure \ref{fig:synth_corr} illustrates the training correlation, the synthetic correlation, and the correlation difference between the original and synthetic training sets. 
As illustrated in the image, the results indicate a mild to strong correlation among the prompt variables: Id, which represents the subject's identity; Description, the prompt description provided in the aforementioned illustration; query, and answer, which denote the query posed to the LLM and the target variable, respectively. The differences in correlation are also minor, except for one subject's query, which suggests an error in data generation from the GAN. 

\subsection{LLM Finetuning and Evaluation}
\label{subsec:res_finetuning}
\begin{figure}[!ht]
    \centering
    \includegraphics[width=1.0\linewidth]{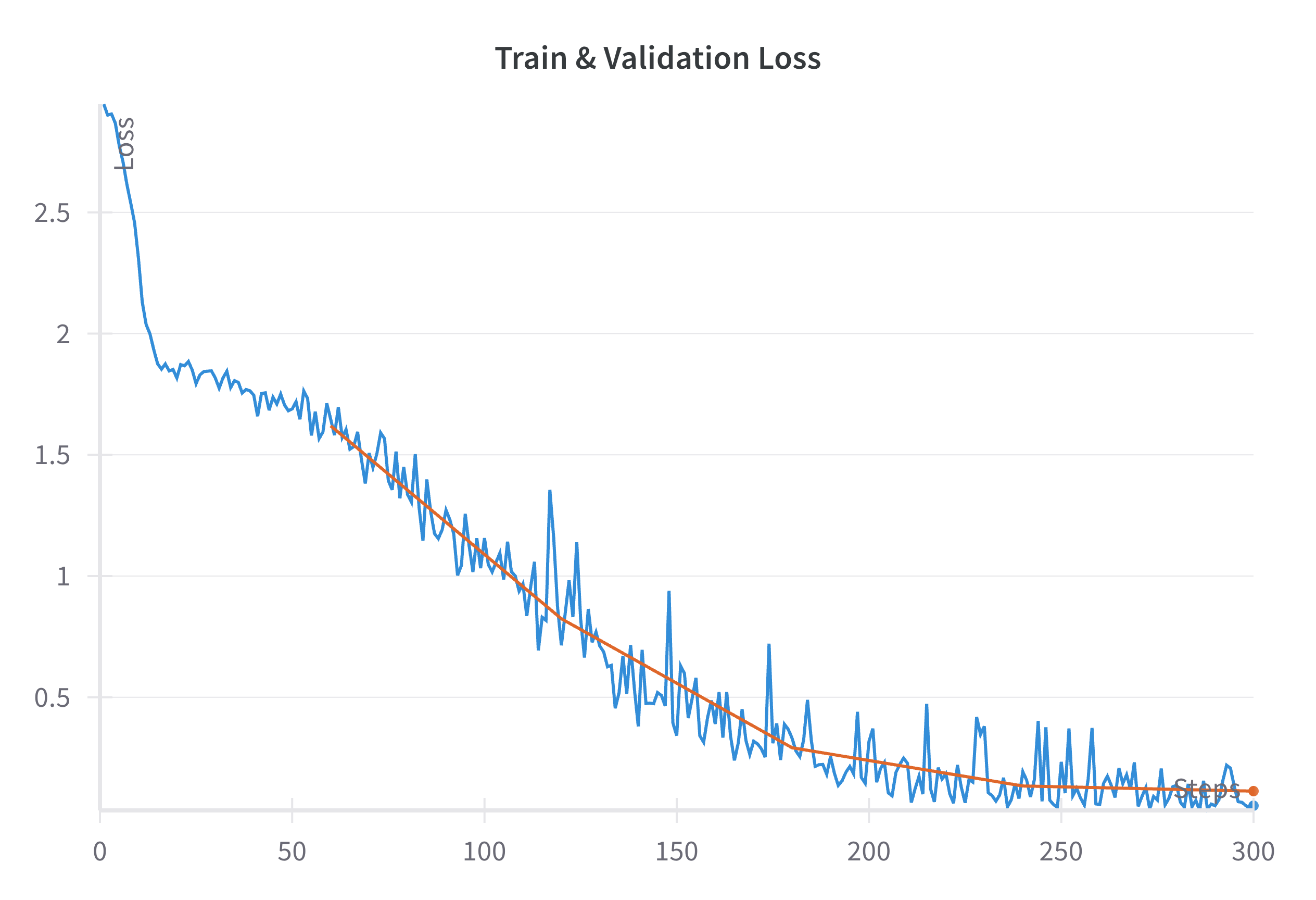}
    \caption{Training and validation loss of LLMs fine-tuning}
    \label{fig:train_eval_loss}
\end{figure}

The fine-tuning and testing processes utilise the Llama 3.1 model, which has 8 billion parameters. The selected model has been vetted against the established considerations outlined in \ref{subsec:fine_tuning}. The Llama models demonstrate strong performance in complex tasks and datasets, and they are well aligned with logical reasoning. The model is available for use without any restrictions. It is worth noting that the author is aware of the many other free options for LLMs available at the time of compiling this manuscript. The open-source nature and free usage were the primary reasons for adopting the Llama 3.1 model. Table \ref{tab:llmparams} provides an overview of the training parameters for the Llama 3.1-8B model.
\begin{table}
    \centering
    \begin{tabular}{|c|c|}
        \hline
        \textbf{LLM Training Parameters} & \textbf{Value} \\ \hline
        Number of trained epochs & 1 \\ \hline
        Gradient Accumulation Steps & 8 \\ \hline
        Optimiser & 32-bit Paged AdamW Optimizer \\ \hline
        Logging steps & 1 \\ \hline
        Weight decay & 0.0002 \\ \hline
        LoRA alpha & 16 \\ \hline
        Max. gradient norm  & 0.3 \\ \hline
        Warmup ratio & 0.03 \\ \hline
        Learning rate scheduler & Cosine \\ \hline
        Evaluation strategy & "steps" by saving checkpoints on every epoch  \\ \hline
        Evaluation steps & 0.2 \\ \hline
    \end{tabular}
    \caption{Training parameters for Llama 3.1 LLM fine-tuning}
    \label{tab:llmparams}
\end{table}
The training process utilised 3,000 prompts for model fine-tuning; computing restrictions did not favour the use of more data in the fine-tuning process. A train-test split strategy of 90/10 was employed, meaning 2,700 prompts were used for training and 300 for testing. Figure \ref{fig:train_eval_loss} depicts the train and evaluation loss of the LLMs fine-tuning.
The LLM model was tested with 300 prompts both before and after fine-tuning, using the same test set. This indicates that the LLM model was evaluated prior to any fine-tuning, and the results were recorded. 
\begin{figure}[!ht]
    \centering
    \includegraphics[width=1.0\linewidth]{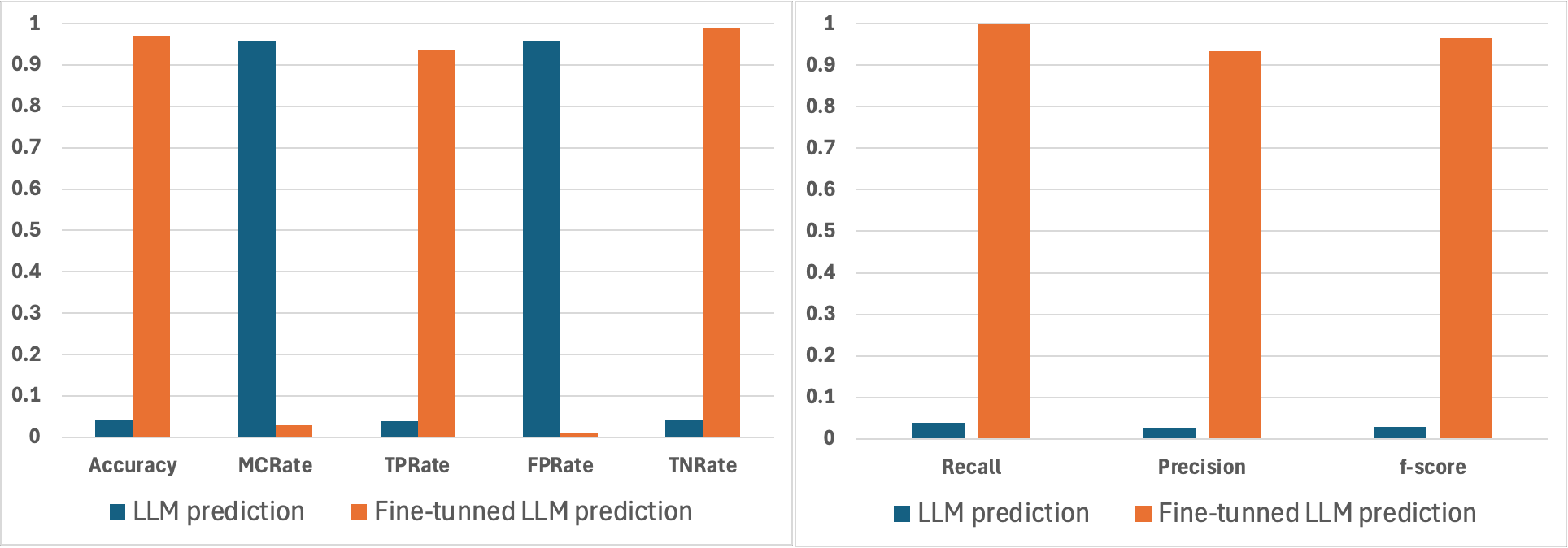}
    \caption{Accuracy, recall, precision and f-score results of LLM performance before and after fine-tuning.}
    \label{fig:res_before_after}
\end{figure}

Figure \ref{fig:res_before_after} illustrates the LLM model's accuracy, misclassification rate (MRate), true positive rate (TRate), false positive rate (FPRate), true negative rate (TNRate), recall, precision, and F-score values. It is clearly evident that LLMs excel when properly fine-tuned with a relatively small amount of specialised and contextualised data. The fine-tuned LLM model was 24 times better than the initial model. 

The initial accuracy of the LLM was considerably low before fine-tuning, achieving only $4.5\%$ compared to $97\%$ after the fine-tuning process. The misclassification rate (MRate) was high at $96\%$ before fine-tuning, in contrast to just $3\%$ following the fine-tuning. Similar improvements were observed in the True Positive Rate (TPRate), which changed from $3.9\%$ to $93.5\%$; the False Positive Rate (FPRate), which dropped from $96\%$ to $1.0\%$; and the True Negative Rate (TNRate), which increased from $4.14\%$ to $98.95\%$. The same improvements can be seen in recall, precision, and f-score measurements, yielding performance metric values of $3.92\%$ and	$99.37\%$ for recall related to before and after the fine-tuning process; $2.40\%$ and	$93.33\%$ for precision; and $2.98\%$	and $96.55\%$ for f-score, respectively.

\section{Conclusion and Future Work}
\label{sec:conclusion}

This paper has addressed the issue of LLM fine-tuning with brain thought processes such as EEG microstates during mental arithmetic tasks. The research contributes to the field of cognitive load studies by introducing the fine-tuning of LLM models with EEG microstate features for detecting cognitive load conditions, including 'Rest' and 'Load' states. It provides a comprehensive and reproducible experimental setup for LLM fine-tuning with EEG microstate data, creating a highly contextualised LLM model. The results demonstrate exceptionally high model performance compared to the same model prior to the proposed fine-tuning. The model's capability to detect cognitive load states increases by approximately 24 times. The direct implications of the study indicate that EEG microstate data can be effectively employed to differentiate between cognitive load conditions. The indirect implications, which necessitate further intrinsic analysis of microstates, suggest that these findings could underpin advancements in our understanding of cognition within the context of AI as a whole and provide a solid foundation for future exploration in the realm of Cognitive AI.

Future opportunities for further research that could advance the initial finding outlined in this research are:
\begin{itemize}
    \item Exploring and comparing the capabilities of other Large Language Models that excel in cognitive tasks, such as Gemini, DeepSeek, Claude, and OpenAI's GPT. 
    \item Designing specialised LLM models that are contextualised for specific cognitive tasks can be implemented in various intelligent agents. For instance, inputting EEG microstate data from a user during particular decision-making tasks, where high alertness or vigilance is necessary (such as driving and operating heavy vehicles) into an LLM model for fine-tuning and assessing the model's decision-making capabilities regarding these decisions. 
    \item Further experiments with different parameters or datasets could enhance the robustness of the findings. We are excited to see how subsequent studies can build upon and extend our work.
\end{itemize}
\subsubsection{Data \& Code Statement: } The data and the code used in this experiment are available at: https://www.kaggle.com/code/braufi/eeg-llama3-finetuning,  upon request from the author. The prompt dataset used in the study is public and available on Huggingface (https://huggingface.co/datasets/braufi/eeg-micostates-promts)
%
%
%

\bibliographystyle{splncs04}
\bibliography{mybibliography}

\begin{thebibliography}{10}
\providecommand{\url}[1]{\texttt{#1}}
\providecommand{\urlprefix}{URL }
\providecommand{\doi}[1]{https://doi.org/#1}

\bibitem{Bréchet2022EEG}
Bréchet, L., Michel, C.: Eeg microstates in altered states of consciousness. Frontiers in Psychology  \textbf{13} (2022). \doi{10.3389/fpsyg.2022.856697}

\bibitem{Chen2023Evidence}
Chen, J., Ke, Y., Ni, G., Liu, S., Ming, D.: Evidence for modulation of eeg microstates by mental workload levels and task types. Human brain mapping  (2023). \doi{10.1002/hbm.26552}

\bibitem{Croce2020EEG}
Croce, P., Quercia, A., Costa, S., Zappasodi, F.: Eeg microstates associated with intra- and inter-subject alpha variability. Scientific Reports  \textbf{10} (2020). \doi{10.1038/s41598-020-58787-w}

\bibitem{Echterhoff2024Cognitive}
Echterhoff, J., Liu, Y., Alessa, A., McAuley, J., He, Z.: Cognitive bias in decision-making with llms. Findings of the Association for Computational Linguistics  (2024)

\bibitem{Goertzel2023Generative}
Goertzel, B.: Generative ai vs. agi: The cognitive strengths and weaknesses of modern llms. ArXiv  \textbf{abs/2309.10371} (2023). \doi{10.48550/arXiv.2309.10371}

\bibitem{Khanna2015Microstates}
Khanna, A.R., Pascual-Leone, A., Michel, C., Farzan, F.: Microstates in resting-state eeg: Current status and future directions. Neuroscience \& Biobehavioral Reviews  \textbf{49},  105--113 (2015). \doi{10.1016/j.neubiorev.2014.12.010}

\bibitem{koenig2002millisecond}
Koenig, T., Prichep, L., Lehmann, D., Sosa, P.V., Braeker, E., Kleinlogel, H., Isenhart, R., John, E.R.: Millisecond by millisecond, year by year: normative eeg microstates and developmental stages. Neuroimage  \textbf{16}(1),  41--48 (2002)

\bibitem{Kojima2022Large}
Kojima, T., Gu, S., Reid, M., Matsuo, Y., Iwasawa, Y.: Large language models are zero-shot reasoners. ArXiv  \textbf{abs/2205.11916} (2022)

\bibitem{lehmann2009eeg}
Lehmann, D., Pascual-Marqui, R.D., Michel, C.: Eeg microstates. Scholarpedia  \textbf{4}(3), ~7632 (2009)

\bibitem{Lian2021Altered}
Lian, H., Li, Y., Li, Y.: Altered eeg microstate dynamics in mild cognitive impairment and alzheimer's disease. Clinical Neurophysiology  \textbf{132},  2861--2869 (2021). \doi{10.1016/j.clinph.2021.08.015}

\bibitem{Mahowald2023Dissociating}
Mahowald, K., Ivanova, A.A., Blank, I., Kanwisher, N., Tenenbaum, J., Fedorenko, E.: Dissociating language and thought in large language models: a cognitive perspective. ArXiv  \textbf{abs/2301.06627} (2023). \doi{10.48550/arXiv.2301.06627}

\bibitem{Metzger2023Functional}
Metzger, M., Dukic, S., McMackin, R., Giglia, E., Mitchell, M., Bista, S., Costello, E., Peelo, C., Tadjine, Y., Sirenko, V., Plaitano, S., Coffey, A., McManus, L., Sharp, A.F., Mehra, P., Heverin, M., Bede, P., Muthuraman, M., Pender, N., Hardiman, O., Nasseroleslami, B.: Functional network dynamics revealed by eeg microstates reflect cognitive decline in amyotrophic lateral sclerosis. Human Brain Mapping  \textbf{45} (2023). \doi{10.1002/hbm.26536}

\bibitem{Michel2017EEG}
Michel, C., Koenig, T.: Eeg microstates as a tool for studying the temporal dynamics of whole-brain neuronal networks: A review. Neuroimage  \textbf{180},  577–593 (2017). \doi{10.1016/j.neuroimage.2017.11.062}

\bibitem{Milz2016The}
Milz, P., Faber, P., Lehmann, D., Koenig, T., Kochi, K., Pascual-Marqui, R.: The functional significance of eeg microstates—associations with modalities of thinking. NeuroImage  \textbf{125},  643--656 (2016). \doi{10.1016/j.neuroimage.2015.08.023}

\bibitem{Momennejad2023Evaluating}
Momennejad, I., Hasanbeig, H., Frujeri, F., Sharma, H., Ness, R., Jojic, N., Palangi, H., Larson, J.: Evaluating cognitive maps and planning in large language models with cogeval. ArXiv  \textbf{abs/2309.15129} (2023). \doi{10.48550/arXiv.2309.15129}

\bibitem{Musaeus2019Microstates}
Musaeus, C., Nielsen, M.S., Høgh, P.: Microstates as disease and progression markers in patients with mild cognitive impairment. Frontiers in Neuroscience  \textbf{13} (2019). \doi{10.3389/fnins.2019.00563}

\bibitem{Nolfi2023On}
Nolfi, S.: On the unexpected abilities of large language models. ArXiv  \textbf{abs/2308.09720} (2023). \doi{10.48550/arXiv.2308.09720}

\bibitem{pascual1995segmentation}
Pascual-Marqui, R.D., Michel, C.M., Lehmann, D.: Segmentation of brain electrical activity into microstates: model estimation and validation. IEEE Transactions on Biomedical Engineering  \textbf{42}(7),  658--665 (1995)

\bibitem{Raiaan2024A}
Raiaan, M.A.K., Mukta, M.S.H., Fatema, K., Fahad, N.M., Sakib, S., Mim, M.M.J., Ahmad, J., Ali, M.E., Azam, S.: A review on large language models: Architectures, applications, taxonomies, open issues and challenges. IEEE Access  \textbf{12},  26839--26874 (2024). \doi{10.1109/ACCESS.2024.3365742}

\bibitem{raufi2022evaluation}
Raufi, B., Longo, L.: An evaluation of the eeg alpha-to-theta and theta-to-alpha band ratios as indexes of mental workload. Frontiers in Neuroinformatics  \textbf{16},  861967 (2022)

\bibitem{Sartori2023Language}
Sartori, G., Orrú, G.: Language models and psychological sciences. Frontiers in Psychology  \textbf{14} (2023). \doi{10.3389/fpsyg.2023.1279317}

\bibitem{Seitzman2017Cognitive}
Seitzman, B., Abell, M., Bartley, S.C., Erickson, M.A., Bolbecker, A., Hetrick, W.: Cognitive manipulation of brain electric microstates. NeuroImage  \textbf{146},  533--543 (2017). \doi{10.1016/j.neuroimage.2016.10.002}

\bibitem{Shaw2019Capturing}
Shaw, S., Dhindsa, K., Reilly, J., Becker, S.: Capturing the forest but missing the trees: Microstates inadequate for characterizing shorter-scale eeg dynamics. Neural Computation  \textbf{31},  2177--2211 (2019). \doi{10.1162/neco_a_01229}

\bibitem{shin2016open}
Shin, J., von L{\"u}hmann, A., Blankertz, B., Kim, D.W., Jeong, J., Hwang, H.J., M{\"u}ller, K.R.: Open access dataset for eeg+ nirs single-trial classification. IEEE Transactions on Neural Systems and Rehabilitation Engineering  \textbf{25}(10),  1735--1745 (2016)

\bibitem{Takarae2022EEG}
Takarae, Y., Zanesco, A., Keehn, B., Chukoskie, L., Müller, R.A., Townsend, J.: Eeg microstates suggest atypical resting-state network activity in high-functioning children and adolescents with autism spectrum development. Developmental science  (2022). \doi{10.1111/desc.13231}

\bibitem{Ville2010EEG}
de~Ville, D.V., Britz, J., Michel, C.: Eeg microstate sequences in healthy humans at rest reveal scale-free dynamics. Proceedings of the National Academy of Sciences  \textbf{107},  18179 -- 18184 (2010). \doi{10.1073/pnas.1007841107}

\bibitem{Zappasodi2019EEG}
Zappasodi, F., Perrucci, M.G., Saggino, A., Croce, P., Mercuri, P., Romanelli, R., Colom, R., Ebisch, S.: Eeg microstates distinguish between cognitive components of fluid reasoning. NeuroImage  \textbf{189},  560--573 (2019). \doi{10.1016/j.neuroimage.2019.01.067}

\bibitem{zhang2024unsupervised}
Zhang, Z.Y., Zhang, J., Yao, H., Niu, G., Sugiyama, M.: On unsupervised prompt learning for classification with black-box language models. arXiv preprint arXiv:2410.03124  (2024)

\bibitem{Zhu2024Language}
Zhu, J.Q., Yan, H., Griffiths, T.L.: Language models trained to do arithmetic predict human risky and intertemporal choice. ArXiv  \textbf{abs/2405.19313} (2024). \doi{10.48550/arXiv.2405.19313}

\bibitem{Zhuang2023Efficiently}
Zhuang, Y., Liu, Q., Ning, Y., Huang, W., Lv, R., Huang, Z., Zhao, G., Zhang, Z., Mao, Q., Wang, S., Chen, E.: Efficiently measuring the cognitive ability of llms: An adaptive testing perspective. ArXiv  \textbf{abs/2306.10512} (2023). \doi{10.48550/arXiv.2306.10512}

\bibitem{zyma2019electroencephalograms}
Zyma, I., Tukaev, S., Seleznov, I., Kiyono, K., Popov, A., Chernykh, M., Shpenkov, O.: Electroencephalograms during mental arithmetic task performance. Data  \textbf{4}(1), ~14 (2019)

\end{thebibliography}
\end{document}